# A transparent radiative cooling photonic structure with a high NIR reflection

Saichao Dang, Hong Ye*

Corresponding Author

Hong Ye-Department of Thermal Science & Energy Engineering, University of Science and Technology of China, Hefei, Anhui 230027, P. R. China. Email: hye@ustc.edu.cn

Authors

Saichao Dang-Department of Thermal Science & Energy Engineering, University of Science and Technology of China, Hefei, Anhui 230027, P. R. China

## Abstract

Buildings or vehicles with transparent envelope can be heated up by sunlight, causing energy consumption for cooling and in extreme cases leading to vehicular heatstroke in a hot climate. Because only visible light for illumination is essential for these applications, the NIR solar radiation should be reflected to reduce heat gain and the infrared radiation emission should be enhanced for further cooling by using the sky as a heat sink. With a high NIR reflection, a transparent radiative cooling photonic structure consisting of 2D silica gratings atop ZnO/Ag/ZnO is demonstrated for energy-saving and safety. With 81% visible light transmitted, 57% NIR solar radiation reflected and 90% thermal infrared radiation emitted, a synthetical cooling is realized by the photonic structure. Theoretically, the total power of reflected solar irradiance and radiative cooling in infrared of this structure is more than double that of a planar silica. The field test shows that with this structure, the temperature rise of a sealed chamber covered by planar silica can be cooled down by 53.7%. This work shows that the concept of daytime radiative cooling can be applied in combination with the utilization of visible light, indicating a great practical application.

**Keywords:** Radiative cooling, photonic structure, D/M/D, NIR reflection

Daytime passive radiative cooling (DPRC) has received great attention in recent years [1-6]. Without any external driving energy input, this cooling method can be extensively applied in energy saving applications such as buildings, photoelectric devices or the thermal management of spacecraft [7-9]. According to the mechanism of this cooling technology, efficient DPRC should have minimum absorption of incoming solar radiation and maximum emission of infrared thermal radiation through the atmospheric window (8~13 $\mu$m) for releasing heat [10]. A classic DPRC material consists of a bottom opaque metallic layer (~150 nm), i.e., Al or Ag [11-16], which can reflect incident sunlight, and a rest top structure which emits in the atmospheric window and doesn't absorb sunlight as well. By tuning the top structure, the developed DPRC materials can be much cooler than the ambient under direct sunlight [1-4]. In recent years, high-performance DPRC has been theoretically and experimentally demonstrated in 1D [4, 5], 2D [6-8], and 3D [9, 14, 16] photonic structures. What's more, to make the DPRC devices economical and manufactured in large scale, polymers and microspheres are applied [13, 17-20]. General DPRC applications should selectively emits thermal radiation in the atmospheric window and minimize solar absorption, simultaneously. However, for a solar cell, the solar radiation at wavelengths with energy higher than the cell bandgap (e.g., 1.1 $\mu$m for crystalline silicon solar cell) is essential. This part of solar radiation should be absorbed by a solar cell to generate power. However, the absorbed energy will in turn increase the cell temperature to 50°C–55°C or even higher [21-25] in practical outdoor conditions, causing significant adverse consequences for the performance and reliability of solar cells. For solar cell cooling, the proposed several photonic structures can strongly radiate heat through the atmospheric window (8−13 $\mu$m) and maintain or even enhance the solar absorption in cell operating band [7, 26-30].

Inspired by the principle of solar cell radiative cooling which requires a transparent band for cell operating, to achieve passive cooling for a transparent envelope on buildings or vehicles in hot climates, only visible light is allowed for illumination. The rest solar radiation (mainly NIR) if transmitted into a building or car through the window will consume an additional energy for cooling. It is effective to improve the energy-saving performance by modifying the radiative properties of windows, i.e., one of the most important envelope components of buildings in which the energy consumption occupies a huge part [4, 30] (reaching 35% in 2020, China [31-34]) among the total national energy consumption. The unwanted solar radiation should be reflected back to outdoor environment. Through radiative cooling in thermal infrared band, the energy consumed by air conditioner will be reduced further. For a vehicle in hot climates, the inside temperature can increase by 40 °C on average over the course of 60 minutes [35, 36]. Due to the rapid temperature rise, as reported by National Safety Council of USA, the total number of U.S. pediatric vehicular heatstroke deaths in 2019 is 52, which could have been prevented [37]. Thus, DPRC for a vehicle is needed for both energy saving and safety. As a common visible transparent NIR reflecting film in energy-saving, a dielectric/metal/dielectric (D/M/D) structure is often used as a spectrally selective filter that can transmit most of the visible light and reflect NIR solar radiation [38-46]. Dielectric layers can enhance visible transmission due to destructive multi-reflective interference and thin metal layer is responsible for NIR reflection [39-42]. However, due to the high reflection of the metal layer, a D/M/D structure shows a low infrared emittance which suppresses the radiative cooling [39-42]. As silica is transparent over most of the solar wavelength range and also exhibits a strong phonon–polariton resonance in the wavelength range of 8~13 $\mu$m, it is expected that two-dimensional silica gratings atop a D/M/D structure can realize a high visible transmission, a high IR reflection and a high infrared emittance, simultaneously.

In this work, a synthetical cooling structure is proposed and demonstrated by a visible-transparent, NIR-reflective radiative cooling photonic structure consisting of 2D silica pillar gratings atop ZnO/Ag/ZnO (SPZAZ). This structure is prepared by magnetron sputtering deposition and photolithography technology. Its radiative properties in solar and thermal infrared band are measured to evaluate its cooling performance. Those properties are also simulated by the rigorous coupled-wave analysis method (RCWA) to study their mechanisms. Finally, the cooling performance of this SPZAZ is investigated theoretically and experimentally.

**Results and discussion**

As shown in Figure 1, the synthetical cooling conception is expected to selectively utilize sunlight and perform radiative cooling in infrared band, simultaneously. The visible light (0.38~0.78 $\mu$m) is allowed to transmit while the NIR (0.78~2.5 $\mu$m) solar radiation are reflected back to the environment. In this work, the ZnO/Ag/ZnO structure is applied to realize high visible transmittance and high NIR reflectance. The top two-dimensional silica pillar gratings are transparent in solar spectrum maintaining the properties of ZnO/Ag/ZnO structure and show a strong emittance in the atmospheric window (8~13 $\mu$m) for radiative cooling.

**The prepared SPZAZ and its radiative properties**

The prepared SPZAZ is exhibited and its morphology features are characterized as shown in Figure 2. With the size of patterned area being 2.5 cm×2.5 cm, the SPZAZ has a high visible transparency (see Figure 2A). Figure 2B shows the 2D periodic pillar gratings with the period of the grating P=7.80 $\mu$m, the diameter and height of a micro $SiO_2$ pillar D=5.40 $\mu$m and t=2.51 $\mu$m, respectively. A small inclined angle of the pillar edge can be observed. According to Figure 2C, the roughness of the top surface of a pillar is 6.2 nm which is small compared with the concerned wavelength. Thus, a smooth surface assumption is reasonable. Bellow the pillar patterns is a silica layer and a triple-layer ZnO/Ag/ZnO film with their thicknesses ($t_2$/$t_3$/$t_4$) being 51.8/14.3/53.9 nm as shown in Figure 2D.

The radiative properties of this SPZAZ in solar and infrared bands are measured to evaluate its cooling performance. As shown in Figure 3A, this SPZAZ shows a low hemispherical reflectance ($R_h$) in UV band. Due to the existence of Ag layer, which has a plasma frequency at ~135 nm [47], the SPZAZ can absorb the UV energy strongly actually [13], protecting the inside stuff from the UV light. Within visible band, the hemispherical transmittance ($T_h$) can reach 90%, indicating a good illumination. The direct transmittance ($T_d$) reaches 72% at λ≈560 nm, meaning a good view through the SPZAZ. This enhanced visible transmittance is caused by the destructive reflective interference of ZnO/Ag/ZnO beneath the silica layer [45, 46, 48-50]. As shown in Figure 3B, due to the middle Ag layer [45, 46], a high reflectance in NIR band is realized, which can even reach 94% at λ=2.16 $\mu$m. In general, high visible transmittance and NIR reflectance are realized by this SPZAZ which are beneficial for illumination and cooling, respectively. The radiative properties of the SPZAZ in solar spectrum show oscillations, with more oscillations at shorter wavelengths. These oscillations don't exist in the radiative properties of a ZnO/Ag/ZnO structure [40, 45, 46]. They are caused by the interference effect of the two parts of the top silica structure in SPZAZ, i.e., beneath silica layer ($t_1$=2.88 $\mu$m) and atop gratings (t+$t_1$=5.39 $\mu$m) (see the mechanism of the solar band properties in S1 in the Supporting information).

For the realization of radiative cooling through the atmospheric window, the emittance of this SPZAZ in thermal infrared band should be enhanced. As shown in Figure 3C, this SPZAZ shows a high emittance within 8~13 $\mu$m, while a 500$\mu$m-thick planar silica (PS) which is the substrate of the SPZAZ has a big valley at 9 $\mu$m due to a strong phonon–polariton resonance. As the RCWA simulation of SPZAZ agrees well with the measured result in this infrared range, through the simulated result, the mechanism of the enhanced emittance can be investigated. According to Kirchhoff's law, the emittance equals the absorbance [47]. Hence, the mechanism of strong absorption can explain the enhanced emittance. According to Ref [51-53], the enhanced absorption of such a grating structure may be caused by the interference effect and the edge diffraction effect, which will be discussed, respectively, as follows.

As silica is an absorbing medium within thermal infrared band, the intensity of the electromagnetic field will attenuate exponentially inside it. The wave will attenuate completely in a thick silica. According to the behavior of electromagnetic wave within an absorbing medium, the penetration depth through which the radiation power is attenuated by a factor of 1/e ($\approx$37%) can be calculated by

$$\delta = \lambda/(4\pi\kappa) \tag{1) [47]}$$

where $\kappa$ is the imaginary part of the complex refractive index, called extinction coefficient. The penetration depth for silica in this thermal infrared band is shown in Figure S3 in the Supporting information. From Figure S3, one can find that the penetration depth is less than the thickness of the silica layer in most range, indicating the interference effect is not significant. A detailed discussion about the interference effect inside silica can be found in S2 in the Supporting information.

Because the wavelength in thermal infrared are comparable to the size of one unit of the pattern, the diffraction into the patterned gratings by edges of the pattern is always taking a part in absorption enhancement according to the Geometrical Theory of Diffraction (GTD) [51, 52]. Taking $\lambda=10$ $\mu$m as an example, the electromagnetic field is concentrated at the edge of the pillar as shown in Figure 3D and 3E, indicating an enhanced absorption. The patterned structure provides redundant edges, which could diffract the incident electromagnetic waves. The diffracted waves will partially enter the patterned structure and thus enhance the overall absorption. For a pyramidal structure, it also utilizes diffraction from wedges and edges to enhance the absorption [54]. Thus, the enhanced emittance in atmospheric window is mainly caused by edge diffraction effect. Figure 3F shows the averaged oblique incident emittance of the SPZAZ within 8~13 $\mu$m. It is calculated by integral of the spectral emittance over the spectral radiation of a 300 K black body. This averaged emittance over both polarizations remains larger than 0.8 between 0° and 40°. What's more, between the incident angle of 0° and 70°, the SPZAZ shows a higher emittance than the PS. The enhanced emittance of this SPZAZ persists to large angles.

**Theoretical cooling performance**

An ideal synthetical cooling approach for a transparent envelope should modulate the solar radiation and thermal radiation perfectly, realizing the most illumination, the least heat gain and the strongest radiative cooling in infrared band. The required radiative properties of an ideal design are shown in Figure 4B and 4C. The solar spectral irradiance at air mass 1.5 ($I_{AM}1.5(\lambda)$) can be calculated as 892.3 W/m² [55]. The visible part (0.38~0.78 μm) carries approximately 52.5% of the entire solar energy and the NIR (0.78~2.5 μm) carries approximately 45.4 %. As the visible light is essential, the reflected power by this ideal design is 892.3×(1-0.525)=423.8 W/m², including a large power of 405.1 W/m² in NIR band (as shown in Table 1) and a small amount power of 18.7 W/m² in UV band.

In solar band, the radiative properties can be defined as the integral of the spectral properties (ξ(λ), reflectance/transmittance) over the solar spectrum weighted by the solar spectral irradiance at $I_{AM}1.5(\lambda)$, viz.,

$$\xi_{\text{solar}} = \frac{\int_{\lambda_1}^{\lambda_2} \xi(\lambda) I_{AM1.5}(\lambda) \mathrm{d}\lambda}{\int_{\lambda_1}^{\lambda_2} I_{AM1.5}(\lambda) \mathrm{d}\lambda} \qquad (2)$$

where $\lambda_1$ and $\lambda_2$ are the concerned wavelength limits. According to this equation, the SPZAZ and a 500μm-thick PS show transmittances of 81% and 93% in visible region, respectively. The SPZAZ has a less heat gain (~56.2 W/m2) in this band. As shown in Table 1, this SPZAZ can reflect 57% NIR solar radiation which is much greater than that reflected by the PS.

Besides reflecting the unwanted NIR energy, the cooling performance can be further realized by enhancing thermal emission in the atmospheric window. To calculate the cooling power through this window, consider the structure of area $A$ at temperature $T$, whose spectral and angular emittance is $\varepsilon(\lambda,\theta)$. In the present study, the high resolution atmosphere transmittance spectrum is taken from Ref [56] as shown in Figure 4C. The structure will emit radiation and at the same time receive radiation coming from the ambient with a temperature of $T_{amb}$ in thermal infrared band. The radiated energy by the structure can be expressed as

$$P_{rad}(T) = A \iint I_{BB}(T,\lambda)\varepsilon(\lambda,\theta)d\lambda \cos\theta d\Omega \tag{3}$$

Here $\int d\Omega = 2\pi \int_0^{\pi/2} \sin\theta d\theta$ is the angular integral over a hemisphere. $I_{BB}(T,\lambda) = \frac{2hc^2}{\lambda^5} \frac{1}{e^{hc/(\lambda k_B T)} - 1}$ is the spectral radiance of a black body at temperature $T$, where h is Planck's constant, $k_B$ is the Boltzmann constant, c is the speed of light. The absorbed power from the incident atmospheric thermal radiation is

$$P_{atm}(T_{amb}) = A \iint I_{BB}(T_{amb},\lambda)\alpha(\lambda,\theta)\varepsilon_{atm}(\lambda,\theta)d\lambda \cos\theta d\Omega \tag{4}$$

By using Kirchhoff's radiation law, structure's absorbance equals its emittance, i.e., $\alpha(\lambda,\theta) = \varepsilon(\lambda,\theta)$. The angle dependent emittance of the atmosphere is given by

$$\varepsilon_{atm}(\lambda,\theta) = 1 - t(\lambda)^{1/\cos\theta} \tag{5}$$

where $t(\lambda)$ is the atmosphere transmittance in the normal direction and $\theta$ is the zenith angle [1]. Assuming $T = T_{amb} = 300$ K, the cooling powers of the three structures in the two atmospheric windows (8~13 μm and 16~25 μm), i.e., $P_{rad}(T) - P_{atm}(T_{amb})$, can be calculated. The cooling powers of SPZAZ and PS within 16~25 μm are obtained according to their emittance in Figure S3. As shown in Table 1, most cooling power is through 8~13 μm window which is investigated by most relevant work [1-14]. The cooling power through 16~25 μm window can reach approximately 40 W/m², which shouldn't be neglected. Generally, the ideal design shows a best cooling performance. As the three structures have strong emission within two atmospheric windows, the difference among their infrared cooling performances is not significant [28, 29]. The cooling performance is dominated by reflecting NIR solar radiation. The power of reflected solar irradiance and radiative cooling in infrared of SPZAZ is more than double that of the PS.

According to an energy balance model as schematically shown in the inset of Figure 4A, another detailed effective radiative cooling power ($P_{e,cool}$) can be calculated by

$$P_{e,cool}(T) = P_{rad}(T) - P_{atm}(T_{amb}) - P_{Sun} - P_{cond+conv} \qquad (6)$$

where $P_{sun}$ is the gained heat from the sun, including the transmitted and absorbed solar radiation. Although the transmitted visible light is essential for illumination, it brings inevitable heat as well. According to Eq. (2), the heat gain from solar radiation can be given by $P_{Sun} = A(1 - R_{solar})$. The power lost due to convection and conduction is

$$P_{cond+conv} = Ah_{com}(T_{atm} - T) \qquad (7)$$

with $h_{com}$ being the combined effective heat transfer coefficient. The effective radiative cooling power ($P_{e,cool}$) with $h_{com}$ =5 or 10 Wm$^{-2}$K$^{-1}$, is calculated and plotted in Figure 4C as a function of structure temperature. Basically, a higher $h_{com}$ causes a higher $P_{e,cool}$ due to the smaller $P_{cond+conv}$. When temperature ($T$) goes up, $P_{e,cool}$ increases as well because a higher temperature means larger $P_{rad}$ and $P_{cond+conv}$. As can be observed, the $P_{e,cool}$ may be negative, meaning the structure is heated rather than cooled. Even for the ideal design, the transmitted solar power is 468.5 W/m$^2$ while the cooling power at $T$=300 K through atmosphere is 192.3 W/m$^2$, indicating an inevitable temperature rise of the structure. The $P_{sun}$ of PS is much larger than that of the SPZAZ and the $P_{rad}$ of PS is slightly smaller than that of the SPZAZ according to Tab. 1. At $T$=300 K, the radiative heat gain (y value) of the PS is 672.9

W/m² while that of the SPZAZ is 437.2 W/m², i.e., 35% less. The SPZAZ shows a better cooling performance than the PS under the same $h_{com}$. By setting $P_{e,\,cool}=0$, the equilibrium temperature can be obtained. With the same $h_{com}$, the equilibrium temperature of the SPZAZ is approximately 12 °C lower than that of the PS, indicating a great cooling performance of the SPZAZ.

**Cooling performance field test**

To estimate the practical cooling performance of this SPZAZ, a field test was demonstrated by two wood chambers with their top sides sealed by the SPZAZ and PS, respectively. With a size of 2.0 cm×2.0 cm×3.7 cm (length×width×depth), each chamber is made of 1.3cm-thick wood. To minimize the heat gain through sunlight, the chambers are wrapped by aluminum foils. This test was conducted at Hefei (117° E, 31° N, China) on 10 November 2020 (a sunny day) as shown in Figure 5A from 11:30 to14:30 with an average global horizontal irradiance of 500 W/m². The ambient temperature and the temperatures of the air inside the chambers are measured by thermocouples (T type), respectively. exposing the samples to the sky

According to the recorded temperatures shown in Figure 5B, the obtained average temperatures of the ambient air, the air sealed with the PS and the air sealed with the SPZAZ are 21.7 °C, 27.3 °C and 24.4 °C, respectively. From the infrared image in Figure 5A, one can also observe that the surface of the PS is hotter than that of the SPZAZ. As the power of reflected solar irradiance and radiative cooling in infrared of the SPZAZ is more than double that of the PS, the supposed temperature rise falls 2.9 °C, i.e., 53.7% of the entire rise. Due to this synthetical cooling structure, an extraordinary cooling performance is realized.

In summary, the concept of daytime passive radiative cooling is used in combination with the utilization of only visible light in buildings or vehicles with transparent envelope for energy-saving and safety. Consisting of 2D silica pillar gratings atop ZnO/Ag/ZnO, a synthetical cooling photonic structure is demonstrated with high visible transmittance, NIR reflectance and thermal infrared emittance in atmospheric window (8~13 $\mu$m), simultaneously. Theoretically, the total cooling power of reflected solar irradiance and radiative cooling in infrared of this structure is more than double that of a planar silica. The field test in a sunny day shows that with this photonic structure the temperature rise of the chamber sealed by planar silica can be cooled down by 53.7%. An extraordinary cooling performance is realized, indicating a great practical application.

## Materials and Method

In this work, the quartz substrate can be assumed as a practical glass which is transparent in the total solar spectrum. On this quartz, the D/M/D film and the 2D pillar gratings were prepared successively as the schematic shown in Figure 1. Triple-layer ZnO/Ag/ZnO was deposited on a 500$\mu$m-thick quartz by magnetron sputtering (Kurt J. Lesker. LAB 18) in a high vacuum chamber with a pressure value of $4.8\times10^{-6}$ torr. Through Plasma Enhanced Chemical Vapor Deposition (PECVD, Oxford, Plasma Pro System100), a 5.39$\mu$m-thick $SiO_2$ layer was deposited atop of ZnO/Ag/ZnO for further photolithography. Then the $SiO_2$ film was coated by a 4$\mu$m-thick photoresist film through spin coater and hot plate (SUSS Lab spin6 and SUSS Delta HP8). The unfinished sample was covered with a Cr etch mask which was prepared by maskless lithograph system (ATD 1500) and exposed by an optical aligner (SUSS MA6). After that, the $SiO_2$ without the coverage of photoresist was etched off to fabricate the grating pattern on the $SiO_2$ film using Reactive Ion Etching (Oxford ICP380). The last step was to remove the remaining photoresist with acetone.

Through a scanning electron microscopy (SEM, SIRION200), the cross-section of the SPZAZ can be observed. The surface roughness of the sample is estimated with an atomic microscope (Bruker, Demension Icon). The radiative properties of the sample under normal incidence in solar radiation band is measured by a UV-visible-NIR spectrophotometers (Agilent Cary 7000 UMS). Both of the direct and hemispherical visible transmittances are obtained. A Fourier Transform Infrared (FTIR) spectrometer (Bruker VERTEX 80) with an external integrating sphere is used to measure the sample's thermal infrared radiative property.

The spectral properties of this SPZAZ is analyzed numerically by a RCWA [47, 57, 58] method which is an effective semi-analytical method that can calculate the radiative properties of a device as a stack of layers. In the simulation model, the normally incident light is assumed to be linearly polarized in x direction. The geometric parameters are taken from the prepared SPZAZ as shown in Figure 2. Based on an improved formulation of scattering matrices [57], the solving process is coded using MATLAB software.

## Acknowledge

This work was funded by the National Natural Science Foundation (No. 51576188). This work was partially carried out at the USTC Center for Micro and Nanoscale Research and Fabrication, and we thank Yu Wei, Yizhao He, Wenjuan Li, Jinlan Peng, and Xiuxia Wang for their help with micro/nano fabrication. The authors also give special thanks to Hongxin Yao and Wei Yang for their help with the measurement of radiative properties. Also, the authors thank Fan Yang for his help with the outdoor test.

# TABLE CAPTIONS

Table 1 The reflected NIR power by the ideal design, SPZAZ and PS as well as their cooling power in two atmospheric windows.

Table 1 The reflected NIR power by the ideal design, SPZAZ and PS as well as their cooling power in two atmospheric windows

| Cooling power (W/m$^2$) | Total | Reflected NIR power | Atmospheric windows | |
| :---: | :---: | :---: | :---: | :---: |
| | | | 8~13 $\mu$m | 16~25 $\mu$m |
| Ideal design | 574.7 | 405.1 | 125.0 | 44.6 |
| Photonic structure (SPZAZ) | 382.5 | 231.3 | 113.6 | 37.6 |
| Planar silica (PS) | 164.9 | 28.4 | 99.3 | 37.2 |

# FIGURE CAPTIONS

Figure 1 Schematic of the synthetical cooling conception. A single unit of the designed SPZAZ is shown on the right side, with the diameter of pillar D, the height of pillar t, the period P, the thickness of $SiO_2$ layer beneath the pillars $t_1$, bottom ZnO/Ag/ZnO layers with their thicknesses $t_2/t_3/t_4$.

Figure 2 The prepared SPZAZ and its morphology features. (A) Photo of the SPZAZ. (B) SEM image of the SPZAZ with t= 2.51 $\mu$m, D=5.40 $\mu$m, $t_1$= 2.88 $\mu$m and P=7.80 $\mu$m. (C) Surface roughness of the top surface of a pillar. (D) Cross-section SEM of the bottom ZnO/Ag/ZnO layer with their thicknesses ($t_2/t_3/t_4$) being 51.8/14.3/53.9 nm.

Figure 3 The measure radiative properties of SPZAZ in UV and visible bands (A) and NIR band (B). (A) The hemispherical reflectance ($R_h$) in UV band, the hemispherical transmittance ($T_h$) and direct transmittance ($T_d$) in visible band. (B) The $R_h$ in NIR band. (C) Emittance of the SPZAZ and PS within 5~13 $\mu$m. At the cross-section of one unit, the electric field distribution (D) and the magnetic field distribution (E) for a wave of λ=10 $\mu$m. (F) Average simulated emittance $\bar{\varepsilon}$ between 8 and 13 $\mu$m of the SPZAZ and PS as a function of polar angle of incidence.

Figure 4 The ideal design for cooling in solar radiation band (A) and thermal infrared band (B). (C) Effective radiative cooling power as a function of structure temperature.

Figure 5 Rooftop test for demonstrating the cooling performance of the SPZAZ. (A) Image of the SPZAZ and PS exposing to the sky with the taken infrared image. The PS shows a higher temperature. (B) Rooftop measurement of the air temperatures in the chamber sealed with the SPZAZ and PS against ambient air temperature on a clear warm day in Hefei (117° E, 31° N, China) (10 November 2020). The air sealed with the PS is on average 5.6 °C hotter than the ambient temperature and the air sealed with the SPZAZ is on average 2.7 °C hotter than the ambient temperature.

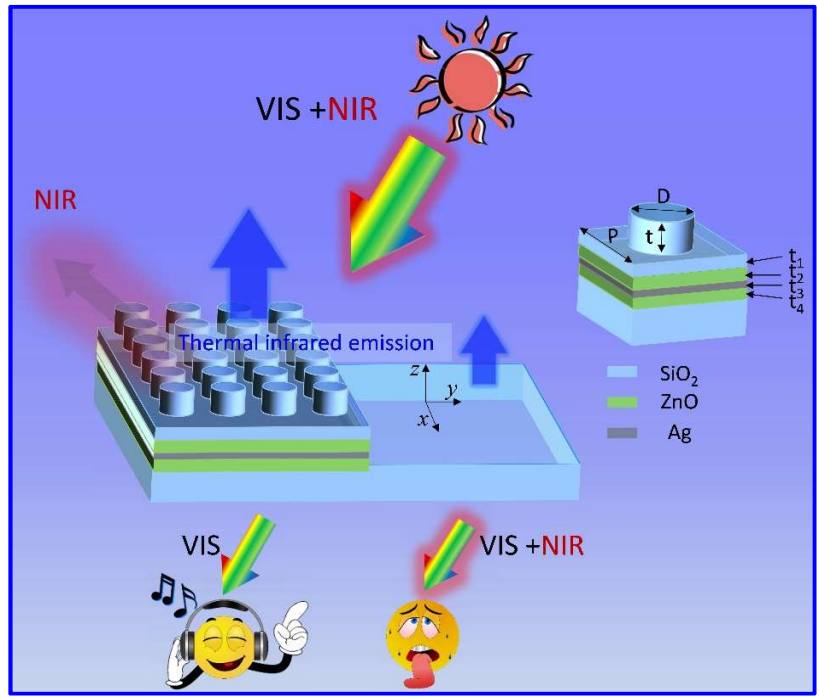

Figure 1 Schematic of the synthetical cooling conception. A single unit of the designed SPZAZ is shown on the right side, with the diameter of pillar D, the height of pillar t, the period P, the thickness of SiO$_2$ layer beneath the pillars $t_1$, bottom ZnO/Ag/ZnO layers with their thicknesses $t_2$/$t_3$/$t_4$.

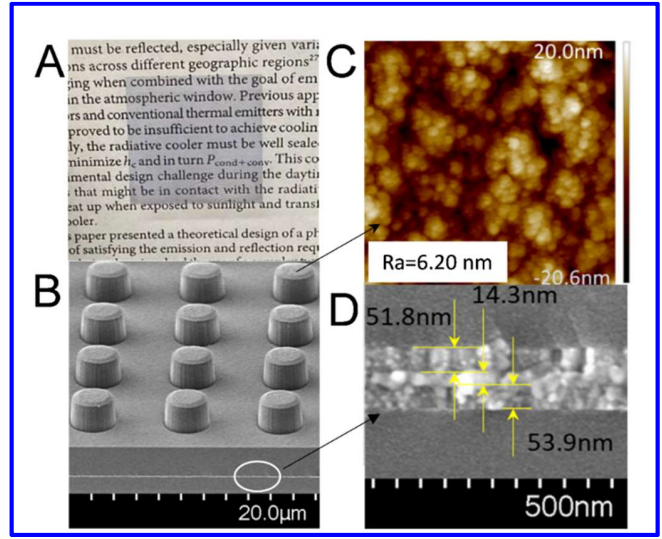

Figure 2 The prepared SPZAZ and its morphology features. (A) Photo of the SPZAZ. (B) SEM image of the SPZAZ with t= 2.51 $\mu$m, D=5.40 $\mu$m, $t_1$= 2.88 $\mu$m and P=7.80 $\mu$m. (C) Surface roughness of the top surface of a pillar. (D) Cross-section SEM of the bottom ZnO/Ag/ZnO layer with their thicknesses ($t_2$/$t_3$/$t_4$) being 51.8/14.3/53.9 nm.

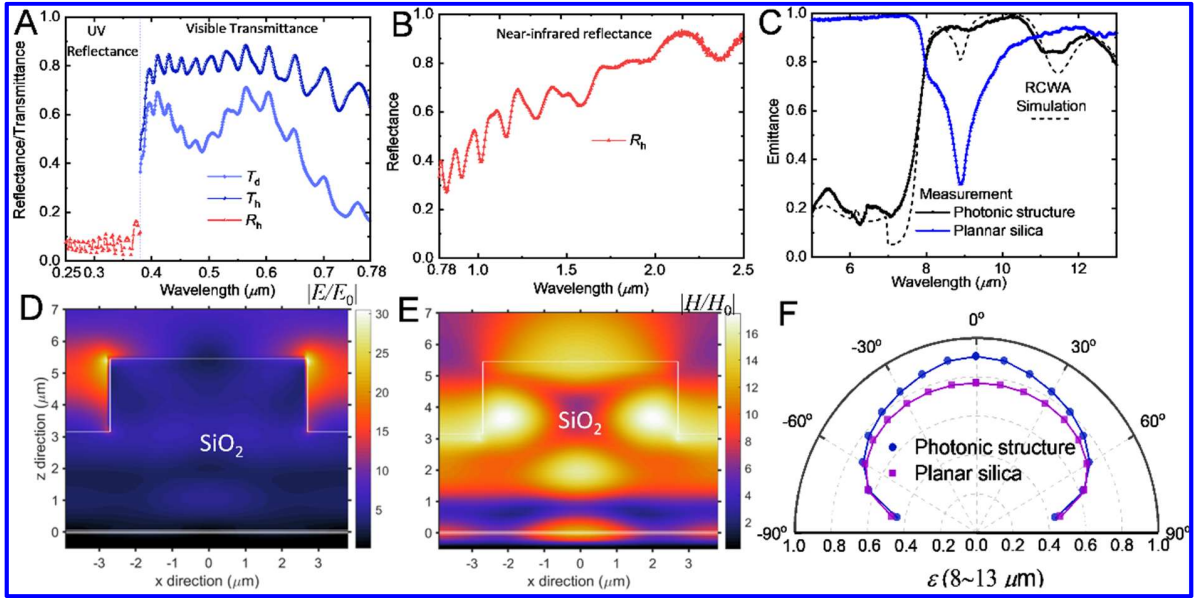

Figure 3 The measure radiative properties of SPZAZ in UV and visible bands (A) and NIR band (B). (A) The hemispherical reflectance ($R_h$) in UV band, the hemispherical transmittance ($T_h$) and direct transmittance ($T_d$) in visible band. (B) The $R_h$ in NIR band. (C) Emittance of the SPZAZ and PS within 5~13 μm. At the cross-section of one unit, the electric field distribution (D) and the magnetic field distribution (E) for a wave of λ=10 μm. (F) Average simulated emittance $\bar{\varepsilon}$ between 8 and 13 μm of the SPZAZ and PS as a function of polar angle of incidence.

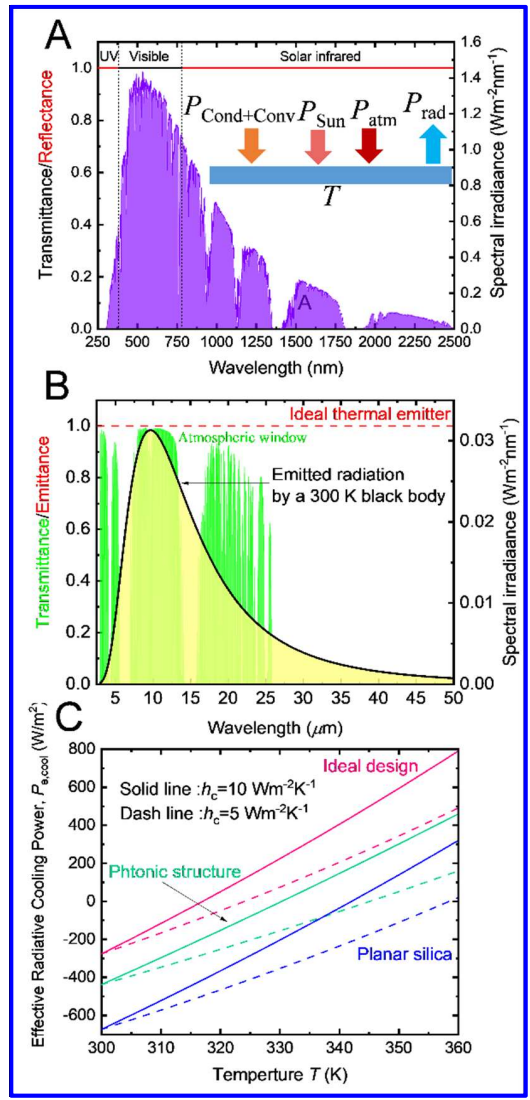

Figure 4 The ideal design for cooling in solar radiation band (A) and thermal infrared band (B). (C) Effective radiative cooling power as a function of structure temperature.

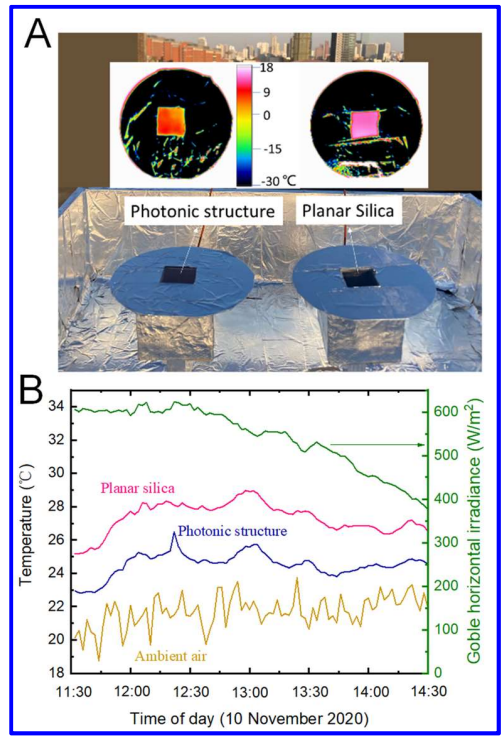

Figure 5 Rooftop test for demonstrating the cooling performance of the SPZAZ. (A) Image of the SPZAZ and PS exposing to the sky with the taken infrared image. The PS shows a higher temperature. (B) Rooftop measurement of the air temperatures in the chamber sealed with the SPZAZ and PS against ambient air temperature on a clear warm day in Hefei (117° E, 31° N, China) (10 November 2020). The air sealed with the PS is on average 5.6 °C hotter than the ambient temperature and the air sealed with the SPZAZ is on average 2.7 °C hotter than the ambient temperature.

29